# Improving Deformable Image Registration Accuracy through a Hybrid Similarity Metric and CycleGAN Based Auto-Segmentation

Keyur D. Shah, PhD[1,2,4], James A. Shackleford, PhD[2], Nagarajan Kandasamy, PhD[2], Gregory C. Sharp, PhD[3]*

[1] Department of Radiation Oncology and Winship Cancer Institute, Emory University, Atlanta, GA 30322
[2] Department of Electrical and Computer Engineering, Drexel University, Philadelphia, PA 19104
[3] Department of Radiation Oncology, Massachusetts General Hospital and Harvard Medical School, Boston 02114
*Email: gcsharp@mgb.org

[4] This work was completed by Keyur D. Shah as a PhD Candidate at Drexel University

# Abstract

**Purpose:** Deformable image registration (DIR) plays a critical role in adaptive radiation therapy (ART) to accommodate anatomical changes. However, conventional intensity-based DIR methods face challenges when registering images with unequal image intensities. In these cases, DIR accuracy can be improved using a hybrid image similarity metric which matches both image intensities and the location of known structures. This study aims to assess DIR accuracy using a hybrid similarity metric and leveraging CycleGAN-based intensity correction and auto-segmentation and comparing performance across three DIR workflows.

**Methods:** The proposed approach incorporates a hybrid image similarity metric combining a point-to-distance (PD) score and intensity similarity score. Synthetic CT (sCT) images were generated using a 2D CycleGAN model trained on unpaired CT and CBCT images, improving soft-tissue contrast in CBCT images. The performance of the approach was evaluated by comparing three DIR workflows: (1) traditional intensity-based (No PD), (2) auto-segmented contours on sCT (CycleGAN PD), and (3) expert manual contours (Expert PD). A 3D U-Net model was then trained on two datasets comprising 56 3D images and validated on 14 independent cases to segment the prostate, bladder, and rectum. DIR accuracy was assessed using Dice Similarity Coefficient (DSC), 95% Hausdorff Distance (HD), and fiducial separation metrics.

**Results:** The hybrid similarity metric significantly improved DIR accuracy. For the prostate, DSC increased from $0.61 \pm 0.18$ (No PD) to $0.82 \pm 0.13$ (CycleGAN PD) and $0.89 \pm 0.05$ (Expert PD), with corresponding reductions in 95% HD from 11.75 mm to 4.86 mm and 3.27 mm, respectively. Fiducial separation was also reduced from 8.95 mm to 4.07 mm (CycleGAN PD) and 4.11 mm (Expert PD) ($p < 0.05$). Improvements in alignment were also observed for the bladder and rectum, highlighting the method's robustness.

**Conclusion:** A hybrid similarity metric that uses CycleGAN-based auto-segmentation presents a promising avenue for advancing DIR accuracy in ART. The study's findings suggest the potential for substantial enhancements in DIR accuracy by combining AI-based image correction and auto-segmentation with classical DIR.

# 1. Introduction

Deformable image registration (DIR) is essential in adaptive radiation therapy (ART), enabling adjustments to treatment plans that account for patient-specific anatomical changes throughout the course of treatment (Oh and Kim 2017, Rigaud et al 2019). These changes can include organ deformation, tumor shrinkage, or other physiological variations, which, if not accounted for, may result in inaccurate dose delivery to the target volume or unintended irradiation of healthy tissues. ART aims to mitigate these risks by allowing treatment adaptation based on pre-treatment imaging, such as cone-beam computed tomography (CBCT).

CBCT imaging is essential in ART workflows due to its ability to provide on-line anatomical information and frequent imaging, enabling regular updates for treatment adaptation (Li et al 2017). However, its lower image quality, particularly in soft tissues (Poludniowski et al 2012), presents significant challenges for accurate DIR (Moteabbed et al 2015, Liu et al 2023). This low contrast, combined with variability in image intensities between planning computed tomography (CT) and CBCT images, results in inaccuracies when relying on intensity-based DIR methods. These methods depend on matching voxel intensities, and often fail when differences in image acquisition, noise, and scatter artifacts obscure anatomical boundaries. Consequently, accurate alignment in regions such as soft tissues becomes difficult to achieve using these techniques.

Conventional DIR techniques, such as cubic B-splines (Rueckert et al 1999, Shackleford et al 2010) and Demons algorithms (Thirion 1998, Muyan-Ozcelik et al 2008, Nithiananthan et al 2009), have long been utilized in clinical practice for registering CT to CBCT images. These approaches attempt to solve the problem of image misalignment by matching voxel intensities between images, often using metrics like the sum of squared differences or mutual information. However, they are sensitive to changes in image quality, such as those introduced by noise or soft-tissue deformation, leading to misalignments in the registration process that can affect the precision of radiotherapy dose delivery. One of the primary limitations of intensity-based DIR is its reliance on global voxel intensity matching. This approach often overlooks fine anatomical details, especially in low-contrast regions, where boundaries between structures are blurred. As a result, important structural information, like the precise location of an organ's surface or tumor margins, can be lost, leading to inaccurate registrations, particularly when significant anatomical changes are present.

To address the issue of poor soft-tissue contrast, researchers have explored incorporating anatomical structures into DIR. By guiding the registration process with known structural information, such as organ contours, these methods improve registration accuracy, especially in regions where intensity-based techniques alone would fail. Early work by Shih et al. (Shih et al 1997) introduced an automated method for anatomic standardization, using contour-based geometric correspondence to refine image registration through non-linear transformations. Lie and Chuang (Lie and Chuang 2003) further advanced this idea by proposing an object-contour-based registration approach, where control points were adaptively selected from contour points to account for both global and local deformations. Gu et al. (Gu et al 2013) introduced a contour-guided DIR (CG-DIR) algorithm that incorporated clinician-edited contours into the registration process, improving the accuracy and consistency of the deformation vector field (DVF) in low-contrast regions.

More recently, efforts have focused on more sophisticated ways of integrating structural information. Zhang et al. (Zhang et al 2023) utilized a combination of surface point clouds and voxel feature points to guide training, achieving significant improvements in registration precision and Lorenzo Polo et al. (Lorenzo Polo et al 2024) introduced a structure-based term into commercial DIR algorithms, resulting in enhanced accuracy in complex pelvic deformations. These studies demonstrated the potential of bio-structure-informed guidance in DIR. These approaches focus on aligning specific regions of interest, such as organs or tumors, rather than relying solely on voxel intensities. By integrating structural information—such as manually delineated contours of key anatomical features, these methods guide the DIR process more effectively, improving accuracy in regions where intensity-based metrics are unreliable.

Another such approach involves the use of a hybrid similarity metric that combines intensity information with structural guidance (Shah et al 2021). The point-to-distance (PD) metric is one such example, where the distance between corresponding anatomical structures on the planning CT and CBCT images is minimized during registration. This allows for a more accurate alignment, particularly in cases where soft tissue contrast is insufficient for intensity-based methods. Studies have demonstrated the efficacy of such approaches, showing enhanced registration performance when structural information is incorporated. However, a key limitation of the PD metric is its reliance on manually segmented contours (Shah 2022), which are time-consuming to create and require expert knowledge. In a clinical setting, manual segmentation can introduce variability and delays, making it impractical for on-line ART workflows.

To overcome the need for manual contours, recent advancements in deep learning (DL) have enabled the automatic segmentation of anatomical structures from medical images. Convolutional neural networks (CNNs), such as the U-Net architecture (Ronneberger *et al* 2015, Isensee *et al* 2021), have shown remarkable success in automatically segmenting structures in CT images where training data is abundant. However, their accuracy is often compromised when applied to low-quality CBCT images due to lack of training data, differences in intensity distribution and overall image quality. To address this limitation, we propose the use of a CycleGAN (Zhu *et al* 2017) for correcting CBCT intensities prior to segmentation. By translating CBCT to sCT with improved soft-tissue contrast, the CycleGAN enables more accurate segmentations, which can then be used to calculate the PD metric without the need for manual input. This approach not only reduces the time required for manual segmentation but also increases consistency across patients, making it a viable solution for ART.

This study builds upon these advancements by integrating DL-based auto-segmentation with the PD metric to improve DIR accuracy. Our proposed method that uses U-Net generated contours as a replacement for manual segmentations in the DIR process, thus eliminating the need for manual input while maintaining high accuracy in the registration process. This novel integration of U-Net generated contours with the hybrid similarity metric presents a practical and scalable approach to enhancing DIR accuracy in ART, particularly in the face of challenging anatomical changes and soft-tissue contrast issues. The approach not only improves the precision of treatment delivery but also streamlines ART workflows, reducing the time clinicians spend on manual segmentation while maintaining consistency across patients. This advancement could potentially lead to more efficient treatment planning and improved patient outcomes in high-volume clinical settings.

## 2. Methods

This study employed deep learning techniques to improve ART workflows by enhancing the quality of CBCT images and automating the segmentation process. A CycleGAN model was used to generate sCT images from CBCT scans, improving soft-tissue contrast and enabling better auto-segmentation. The U-Net model was then applied to segment organs of interest, and these auto-segmentations were used to calculate the PD metric, which refined the DIR process. Imaging data for the study was sourced from the publicly available Pelvic Reference Dataset (PRD) (Yorke et al 2019) and internal clinical data from Massachusetts General Hospital (MGH). In the following sections, we provide details on the patient cohorts, image processing pipeline, and evaluation metrics used to assess the performance of the hybrid registration approach.

2.1 Patient Cohort

The PRD dataset consisted of CT and CBCT scans from 58 patients (both male and female) treated with pelvic radiotherapy. This dataset formed the foundation for training and validating the hybrid deformable image registration approach. Among these, prostate, bladder, and rectum were manually segmented by a clinical expert (GCS) for PRD images with an intact prostate (n=12). The MGH clinical data included CT and CBCT scans from 9 male patients, all with an intact prostate. Additionally, CT scans from 28 other male patients were included in the dataset. Prostate, bladder, and rectum were also manually segmented by the same clinical expert (GCS) for the MGH patients, ensuring consistency across all datasets. Figure 1 outlines the patient flow for both training and validation across the datasets used in this study.

Fiducial markers, which are routinely placed in the prostate for clinical localization, were used as an additional tool to validate DIR accuracy across both the PRD and MGH cohorts. For each prostate, three fiducial markers were identified, and the top and bottom of each fiducial were marked on both CT and CBCT images. The Euclidean distance between corresponding fiducial points on the fixed and registered images was computed after DIR, providing an independent measure of registration accuracy.

For the CycleGAN model, 46 patients from the PRD dataset (including males without an intact prostate and females) were used for training, while the remaining 12 patients were allocated for validation. Additionally, 9 CT and corresponding CBCT images from the internal MGH dataset were used to generate synthetic images (sCT), contributing to the validation of the CycleGAN.

For the U-Net auto-segmentation, a total of 70 images were used, comprising 28 CT-only datasets and 42 paired CT and sCT images (from both PRD and MGH data). The 21 paired datasets included manual contours for the CT images, while the sCT images used contours transferred from CBCT for consistency. An 80-20 train-test split was applied, with 56 images used for training and 14 images (comprising 7 CT and 7 sCT images from the same patients) used for independent testing. This balanced approach ensured that the test set represented both sCT and CT images, allowing robust validation of the segmentation model. The same 7 CT and 7 sCT images, drawn from both PRD and MGH datasets, were also utilized for DIR validation, enabling consistent evaluation of the hybrid registration approach across both segmentation and DIR accuracy metrics. All patient images were anonymized in accordance with institutional review board (IRB) protocols. Inclusion criteria required each patient to have at least one planning CT scan and one or more CBCT scans.

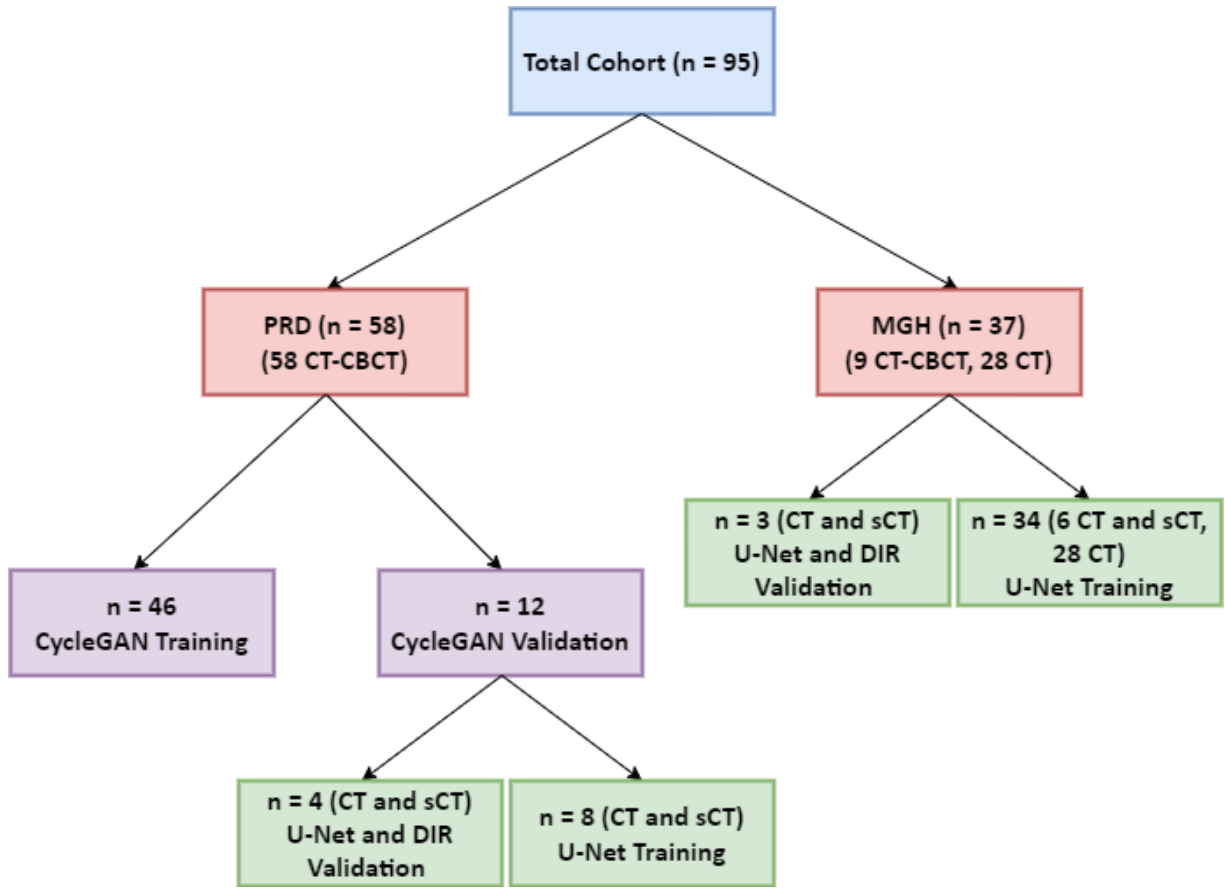

**Figure 1** CONSORT diagram illustrating the allocation of patient datasets across the study workflow. (PRD: Pelvic Reference Dataset; CT: Computed Tomography, sCT: synthetic CT)

2.2 Hybrid Similarity Metric

Our approach employed a hybrid similarity metric that combined intensity-based similarity with the PD Metric to guide the DIR process. The PD metric introduced by Shah et al. (Shah *et al* 2021), developed as part of the Plastimatch (Sharp et al 2010) toolkit (an open-source software suite for medical image analysis) enhanced registration accuracy by penalizing the total distance between corresponding anatomical structures in the fixed and moving images. The metric included a term that minimized the distance between points on a manually or automatically segmented structure boundary (e.g., prostate or bladder) on the fixed (CT) image, and the corresponding boundary in the moving (CBCT) image, using an unsigned distance map.

Formally, the PD metric for a given structure is computed as follows:

$$C = \sum_{T(\theta) \in \Omega} \Psi(f, m) + \lambda S(v) + \sum_{n=1}^{N} \eta_n \sum_{\pi_n \in \Pi_n} |dmap(\pi'_n)| \quad (1)$$

where $\Psi(f, m)$ is the intensity-based similarity metric, $\lambda$ is the regularization penalty term weight applied to $S(v)$ curvature regularizer to smoothen the deformation field, $\eta_n$ is the weight applied to the PD metric for structure $n$, and $dmap(\pi'_n)$ is the unsigned distance map at point $\pi'_n$ is the moving image. This ensured that structural information guides the registration process, improving accuracy and robustness in challenging regions where intensity-based methods may fail.

In our study, the PD metric was applied to prostate, bladder, and rectum contours. The PD metric was optimized through a quasi-Newton optimization algorithm (L-BFGS-B) to minimize both intensity dissimilarity and structural misalignment, yielding a more accurate registration outcome compared to intensity-only methods.

2.3 CycleGAN for Synthetic CT Generation

We employed a CycleGAN model (Zhu et al 2017) to translate CBCT images into sCT images. CycleGAN is particularly well-suited for this task due to its ability to perform unpaired image-to-image translation, thus removing the need for strictly paired CT-CBCT datasets. The model was implemented as a 2D architecture, processing individual axial slices independently to simplify training and ensure compatibility with the available dataset. The CycleGAN architecture consists of two generators and two discriminators. One generator translates CBCT images into sCT images, while the reverse generator translates CT images back into CBCT-like images. The corresponding discriminators aim to distinguish real from synthetic images in both directions, forcing the generators to produce realistic output. To maintain anatomical consistency, we employed a cycle-consistency loss that ensured that an image translated from CBCT to sCT and back to CBCT resembles the original input image. Additionally, an identity loss was employed to preserve anatomical features in the target domain, ensuring that sCT images retain the key anatomical structures necessary for accurate segmentation and DIR.

Axial slices were extracted from the CT and CBCT volumes in DICOM format. To ensure balance between the two sets of images, axial slices encompassing the lower pelvis to the sacral region were selected from the CT and CBCT volumes. This range included critical structures such as the prostate, bladder, and rectum, providing comprehensive coverage for training and validation. After extraction, voxel Hounsfield Unit (HU) values for both CT and CBCT images were truncated at 900 to mitigate the effect of high-intensity metal. The images were then converted to PNG format, intensities were normalized to the range [-1, 1] and resized to 256 × 256 for input into the CycleGAN model.

The CycleGAN model was implemented using Keras with TensorFlow as the backend and trained on a system with an NVIDIA GeForce GTX 1080 Ti (12 GB). The model was optimized using the ADAM optimizer with a learning rate of $10^{-4}$ and $\beta_1 = 0.5$, with a total of 30 epochs. The loss function was a combination of cycle consistency loss ($\lambda_{cycle} = 100$) and identity loss ($\lambda_{identity} = 50$) to regularize the generator networks. The performance of the CycleGAN was assessed visually after each epoch, focusing on the quality of the generated sCT images. Epoch 28 was selected as the final model for image-to-image translation. After this point, the generator began to exhibit mode collapse, producing output images that exhibited memorization.

2.4 U-Net Auto-Segmentation

To automate the segmentation of anatomical structures required for DIR, we utilized a 3D U-Net architecture (Çiçek *et al* 2016). The U-Net model, originally introduced for biomedical image segmentation, consists of a contracting path (encoder) and an expansive path (decoder), with skip connections between layers to preserve spatial information. Our model was adapted to segment volumetric data (3D images) and was trained on the sCT images generated by CycleGAN as well as manually contoured CT images from the PRD and MGH datasets. The U-Net model was trained to segment the prostate, bladder, and rectum. Since the sCT images provide better soft-tissue contrast compared to raw CBCT images, training the U-Net on this dataset greatly improved segmentation accuracy, particularly for challenging regions like the prostate-rectum interface. Data augmentation techniques, including random rotations, translations, and intensity scaling, were applied during training to improve model generalizability. Segmentation performance was evaluated on a holdout set of images, and performance was measured using the Dice Similarity Coefficient (DSC).

The U-Net architecture was implemented and trained using the MONAI framework, a PyTorch-based framework specifically designed for deep learning in healthcare imaging. Training was conducted on a system with an NVIDIA GeForce GTX 1080 Ti (12 GB). The model was optimized using the ADAM optimizer with a learning rate of $10^{-4}$ and trained for 500 epochs. Model performance was monitored after each epoch by assessing both the average training loss and the average validation DSC. The model weights from epoch 481 (which corresponded to the highest validation DSC) were selected for the final auto-segmentation task. The final trained model was used to automatically segment the prostate, bladder, and rectum in both sCT and CT images. These segmented structures were then employed to compute the PD metric for deformable image registration.

2.5 DIR Evaluation

To assess the accuracy of the hybrid DIR process, we employed several evaluation metrics. DSC was used to quantify the overlap between segmented structures on the fixed and moving images. Higher DSC values indicate better registration accuracy. Additionally, the 95% Hausdorff Distance (HD) was calculated to measure the maximum distance between the boundaries of the fixed and registered structures. The Euclidean distance was computed between the marked fiducial points on the fixed and moving images after registration. This serves as an independent measure of registration accuracy, as fiducial alignment is critical for ensuring precise dose delivery in radiotherapy.

To evaluate the DIR methods, three distinct groups were analyzed:
- No PD: Traditional intensity-based DIR approach, without the PD metric.
- CycleGAN PD: The 3D U-Net model was used to segment sCT images, and the PD metric is used during registration.
- Expert PD: Expert manual segmentations were used with the PD metric, serving as a benchmark for comparison with the automated methods.

2.6 Modified DIR Workflow

In standard workflows, CBCT images are used directly for DIR, with conventional intensity-based methods. This study introduced a modified DIR workflow to improve accuracy using deep learning techniques. sCT images generated by CycleGAN are segmented with a 3D-Unet model, and the resulting structures guide the DIR process through the PD metric, enhancing alignment. The overall modified workflow is illustrated in Figure 2, showcasing the integration of CycleGAN-generated sCT, U-Net-based segmentation, and PD metric-guided B-spline registration.

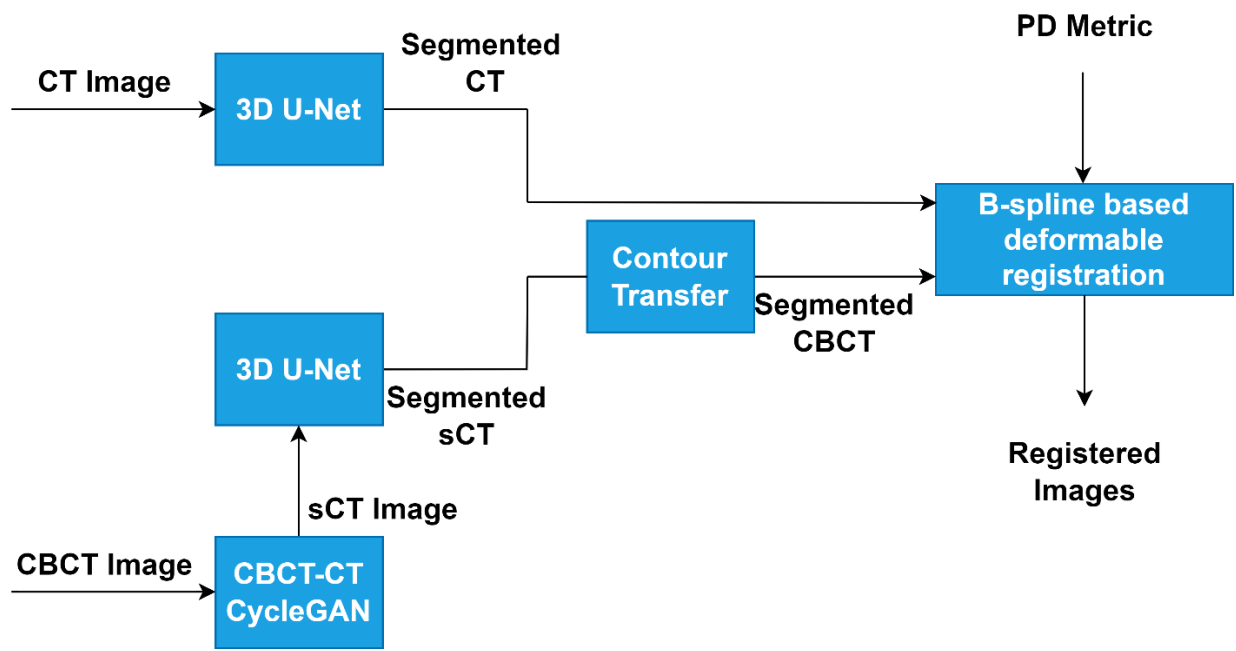

**Figure 2** Overview of the modified DIR workflow integrating CycleGAN-based sCT generation, U-Net segmentation, and PD metric-guided registration for improved accuracy.

2.7 Statistical Analysis

The normality of all data was assessed using the Shapiro-Wilk test. For normally distributed data, one-way ANOVA was used to compare the DSC, 95% HD, and fiducial separation across the three groups (No PD, CycleGAN PD, and Expert PD). When the assumption of normality was violated, the non-parametric Kruskal-Wallis test was employed instead. Post-hoc pairwise comparisons were performed using Tukey's Honest Significant Difference (HSD) test for ANOVA or Bonferroni correction for the Kruskal-Wallis test. In cases where only two groups were compared (e.g., sCT vs. CT for DSC and 95% HD), paired t-tests were applied to normally distributed data, and the Wilcoxon signed-rank test was used for non-normally distributed data. A p-value of less than 0.05 was considered statistically significant for all tests. All statistical analyses were conducted using Python (version 3.9.11) and SciPy (version 1.8.1).

## 3 Results

3.1 CycleGAN Image Quality Improvement

The CycleGAN model greatly improved the soft-tissue contrast in CBCT images. A qualitative comparison between the original CBCT image and the corresponding sCT image for a representative patient is presented in Figure 3. This improvement in soft-tissue contrast was particularly evident in the regions near the prostate, bladder, and rectum.

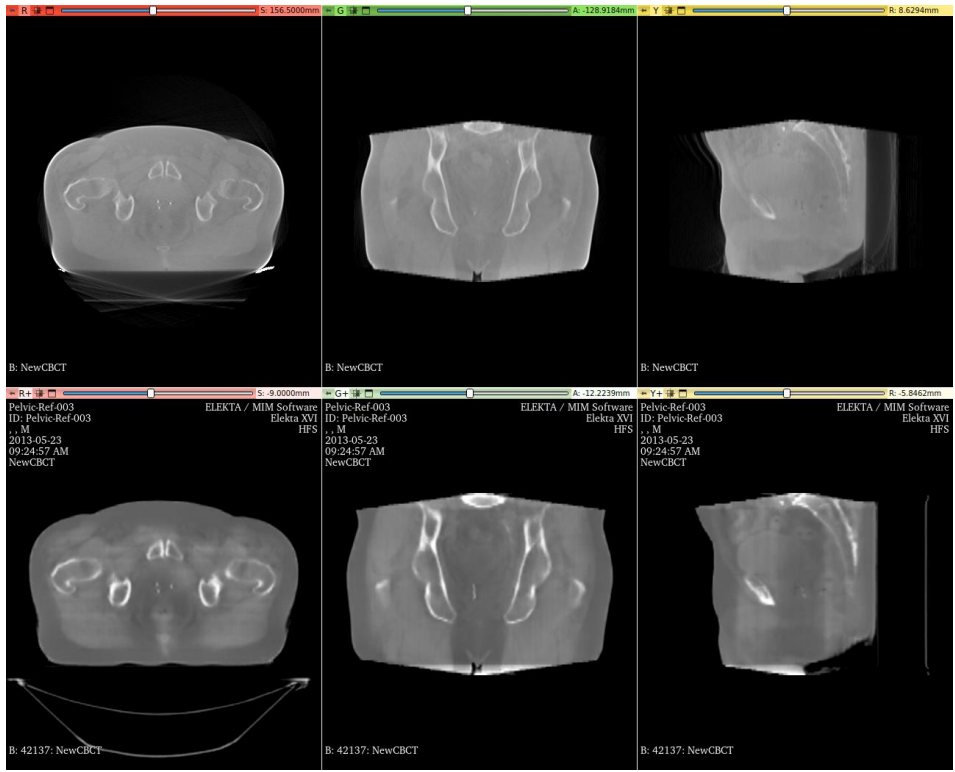

**Figure 3** Qualitative comparison of CBCT (top row) and synthetic CT (sCT) images (bottom row) generated using CycleGAN.

3.2 Auto-Segmentation Results

The 3D U-Net architecture demonstrated robust segmentation performance on both sCT images (generated using CycleGAN) and original CT images. This comparison assessed whether segmentations derived from sCT images were of similar quality to those derived from CT images, highlighting the viability of the CycleGAN-corrected sCT for DIR workflows. Table 1 summarizes the DSC and 95% HD metrics for key anatomical structures across the two modalities.

Table 1: Segmentation Results (DSC and 95% HD) for sCT and CT Images

| Organ | DSC (sCT) (Mean ± SD) | DSC (CT) (Mean ± SD) | p-value (DSC) | 95% HD (sCT) (Mean ± SD) | 95% HD (CT) (Mean ± SD) | p-value (95% HD) |
|---|---|---|---|---|---|---|
| Prostate | 0.80 ± 0.03 | 0.80 ± 0.08 | 0.841 | 3.19 ± 0.87 | 3.22 ± 1.02 | 0.928 |
| Bladder | 0.87 ± 0.06 | 0.89 ± 0.06 | 0.297 | 3.10 ± 1.91 | 5.03 ± 5.71 | 0.917 |
| Rectum | 0.76 ± 0.08 | 0.82 ± 0.09 | 0.024 | 5.13 ± 2.10 | 4.05 ± 3.36 | 0.446 |

For the prostate and bladder, segmentation performance was comparable between sCT and CT images, with no statistically significant differences in DSC or 95% HD. Notably, the rectum segmentations showed a statistically significant difference in DSC ($p < 0.05$), with CT outperforming sCT. However, this difference in rectum segmentation had a minimal effect on boundary delineation, as measured by 95% HD. These results highlight that CycleGAN-generated sCT images enable segmentation accuracy comparable to CT for most structures. Figure 4 is an example of U-Net generated segmentation on a CycleGAN generated sCT image.

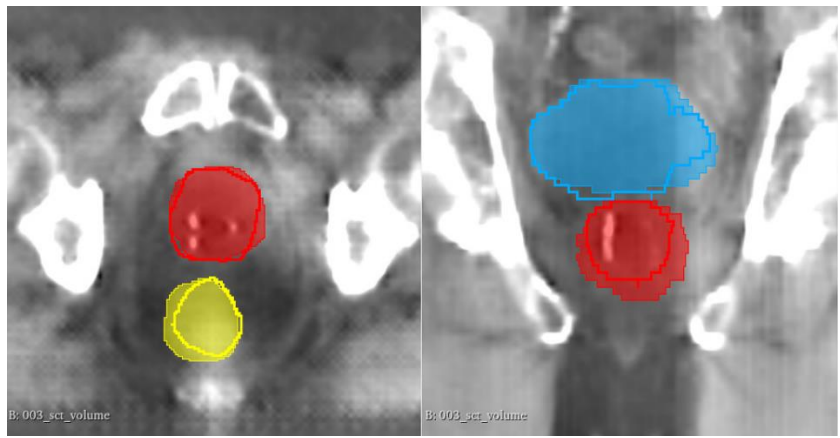

**Figure 4** Example of U-Net segmentation on synthetic CT (sCT) image, illustrating contours for the prostate (red), bladder (blue), and rectum (yellow). Filled indicates ground truth and hollow indicates U-Net generated structures.

3.3 DIR Accuracy Improvement

The hybrid similarity metric led to statistically significant improvements in DIR accuracy across all structures in the validation set. Overall, percentage improvements in DSC and 95% HD ranged from a 56.64% increase in DSC for the bladder to a 73.42% reduction in 95% HD, highlighting the benefit of incorporating the hybrid similarity metric. For the prostate, the inclusion of the PD metric resulted in an increase in DSC from 0.61 ± 0.15 (No PD) to 0.82 ± 0.10 (CycleGAN PD), compared with 0.89 ± 0.05 (Expert PD). There were corresponding reductions in the 95% HD from 11.75 ± 4.93 mm (No PD) to 4.86 ± 2.05 mm (CycleGAN PD), compared with 3.27 ± 0.43 mm (Expert PD) (Table 2). These improvements were statistically significant ($p < 0.05$) between the Expert PD and No PD groups, as well as between CycleGAN PD and No PD groups.

For the rectum, DSC improved from 0.57 ± 0.12 (No PD) to 0.83 ± 0.09 (CycleGAN PD) and 0.89 ± 0.05 (Expert PD), while the 95% HD reduced from 13.58 ± 6.25 mm (No PD) to 5.82 ± 2.21 mm (CycleGAN PD) and 3.61 ± 1.01 mm (Expert PD). Similarly, the bladder exhibited DSC increases from 0.62 ± 0.10 (No PD) to 0.85 ± 0.06 (CycleGAN PD) and 0.94 ± 0.02 (Expert PD). The corresponding 95% HD decreased from 19.59 ± 21.03 mm (No PD) to 21.48 ± 23.90 mm (CycleGAN PD) and 3.09 ± 0.69 mm (Expert PD). However, the 95% HD improvement for the bladder with CycleGAN PD was not statistically significant ($p > 0.05$). Analysis of fiducial markers further validated these improvements. For the prostate, fiducial separation decreased from 8.95 ± 4.02 mm (No PD) to 4.07 ± 1.48 mm (CycleGAN PD) and 4.11 ± 0.82 mm (Expert PD), highlighting enhanced alignment when using the hybrid similarity metric ($p < 0.05$).

**Table 2** Deformable Image Registration (DIR) accuracy results for the prostate, bladder, and rectum, evaluated on the validation cohort, comparing manual expert segmentations (Expert PD), CycleGAN PD (Cycle PD), and no PD metric (No PD). Results include Dice Similarity Coefficient (DSC), 95% Hausdorff Distance (HD), and fiducial separation. Statistical significance was assessed via ANOVA and post-hoc Tukey's HSD.

| Organ | Metric | Expert PD (Mean ± SD) | CycleGAN PD (Mean ± SD) | No PD (Mean ± SD) | p-value (ANOVA) | p-value (Tukey HSD) |
|---|---|---|---|---|---|---|
| Prostate | DSC | 0.89 ± 0.05 | 0.82 ± 0.13 | 0.61 ± 0.18 | 0.002 | Cycle PD vs. No PD: 0.015<br>Expert PD vs. No PD: 0.002 |
| | 95 % HD (mm) | 3.27 ± 0.39 | 4.86 ± 2.48 | 11.75 ± 5.30 | <0.001 | Cycle PD vs. No PD: 0.004<br>Expert PD vs. No PD: <0.001 |
| | Fiducial Separation (mm) | 4.11 ± 0.85 | 4.07 ± 1.20 | 8.95 ± 5.70 | 0.007 | Cycle PD vs. No PD: 0.014 |

| | | | | | | Expert PD vs. No PD: 0.015 |
|---|---|---|---|---|---|---|
| Bladder | DSC | 0.94 ± 0.02 | 0.85 ± 0.11 | 0.62 ± 0.15 | <0.001 | Cycle PD vs. No PD: 0.002<br>Expert PD vs. No PD: <0.001 |
| | 95 % HD (mm) | 3.09 ± 0.62 | 21.48 ± 27.16 | 19.59 ± 15.68 | 0.1414 | Kruskal-Wallis: 0.005 |
| Rectum | DSC | 0.89 ± 0.04 | 0.83 ± 0.08 | 0.57 ± 0.19 | <0.001 | Cycle PD vs. No PD: 0.002<br>Expert PD vs. No PD: <0.001 |
| | 95 % HD (mm) | 3.61 ± 1.03 | 5.82 ± 2.63 | 13.58 ± 5.40 | <0.001 | Cycle PD vs. No PD: 0.001<br>Expert PD vs. No PD: <0.001 |

3.4 Statistical Significance

As shown in Table 2, the statistical significance of the results was evaluated using ANOVA and post-hoc Tukey's HSD test for pairwise comparisons. Significant statistical differences ($p < 0.05$) were observed in the DSC, 95% HD, and fiducial separation metrics across the prostate, bladder, and rectum. For the prostate, statistically significant improvements ($p < 0.05$) were found between the CycleGAN PD and No PD groups, as well as between the Expert PD and No PD groups for both DSC and 95% HD. Fiducial separation within the prostate, a critical target for radiotherapy, was significantly reduced ($p < 0.05$) in both the CycleGAN PD and Expert PD groups compared to No PD group. This result underscores the enhanced alignment accuracy achieved by incorporating the PD metric.

For the rectum, similar statistically significant improvements ($p < 0.05$) were observed for both DSC and 95% HD between the Auto-segmentation and No PD groups, as well as between the Expert PD and No PD groups. Although the bladder demonstrated improvements in DSC, the 95% HD values did not achieve statistical significance for the CycleGAN PD group ($p > 0.05$). Detailed statistical results for the prostate, bladder, and rectum, including DSC, 95% HD, and fiducial separation metrics, are provided in Table 2.

## 4. Discussion

This study demonstrated the integration of a CycleGAN-generated sCT workflow into a hybrid similarity metric for DIR, leveraging intensity-based metrics alongside a structure-guided PD metric. The novelty of this approach lies in the use of CycleGAN to enhance CBCT images, enabling more accurate automatic segmentation via a 3D U-Net model. These improved segmentations, in turn, serve as inputs for the PD metric optimization, significantly enhancing DIR accuracy. The results show statistically significant improvements ($p < 0.05$) in DSC and 95% HD across key anatomical structures, including the prostate, bladder, and rectum, compared to traditional intensity-based DIR methods. Fiducial-based evaluations further validated these accuracy improvements, with significant reductions in fiducial separation post-registration ($p < 0.05$).

Traditional intensity-based DIR methods face well-documented challenges, particularly in low-contrast scenarios such as CBCT imaging. Previous studies like Rivest-Hénault et al. (Rivest-Hénault et al 2014) have incorporated contour-based metrics to address these limitations. Our results demonstrated that combining these metrics with CycleGAN-enhanced segmentation significantly improves registration accuracy. The incorporation of the PD metric in conjunction with AI-generated contours effectively bridges the gap in scenarios where traditional methods fail, particularly in delineating soft-tissue structures. This highlights the potential of advanced deep learning workflows in refining ART processes.

CycleGAN-generated sCT images effectively enhance soft-tissue contrast in CBCT imaging workflows. Unlike previous approaches relying on paired CT-CBCT datasets (Harms et al 2019, Chen et al 2021), our approach utilizes unpaired image translation as demonstrated by Kurz et al. (Kurz et al 2019) and Gao et al. (Gao et al 2021), offering greater flexibility in dataset availability while serving as a key step for accurate DIR. The improved soft-tissue contrast in sCT images not only facilitates segmentation accuracy but also directly impacts the reliability of the PD metric, which relies on accurate contours for optimization. The U-Net trained on sCT images demonstrated segmentation improvements consistent with prior studies (Hesamian et al 2021, Dai et al 2021) and outperformed methods trained on raw CBCT images due to the superior image quality of sCT. These findings underscore the interconnected nature of segmentation and DIR accuracy in workflows reliant on CBCT imaging.

Integrating structural guidance with deep learning-based segmentation, our hybrid approach addresses key challenges in low-contrast CBCT-based DIR while maintaining clinical interpretability. Unlike fully automated deep learning DIR methods, which may face challenges in generalizability and regulatory acceptance, our method balances innovation with practicality, making it more immediately implementable in ART workflows. The improvements demonstrated in registration accuracy highlight the potential for reducing safety margins and improving dose delivery precision, underscoring its clinical relevance.

This study demonstrates that integrating the PD metric into the DIR process enables more precise image registration, particularly for low-contrast CBCT images. The statistically significant reductions in fiducial separation highlight the clinical potential for improving alignment accuracy in prostate-targeted ART. These results suggest the possibility of reducing safety margins, thereby sparing healthy tissues from unnecessary radiation exposure. The workflow's reliance on automatic segmentation from CycleGAN-generated sCT images addresses critical bottlenecks in ART, particularly in high-volume clinics. By reducing the need for manual contouring, this approach not only streamlines treatment planning but also paves the way for integrating automated, on-line ART workflows into clinical practice.

Despite these promising results, several limitations must be considered. First, the accuracy of the DIR process using the PD metric is heavily influenced by the quality of the input segmentations. The results demonstrated in this study are contingent on the accuracy of the U-Net auto-segmentation model. Poor segmentation quality can propagate through the registration process, particularly in anatomically complex regions. As a result, the PD metric may underperform in cases where segmentation accuracy is insufficient.

The reliance on large, high-quality datasets to train the CycleGAN and U-Net models could also present a limitation. While our results were promising in our datasets of 95 patients, the generalizability of the models to other anatomical sites or patient populations may require more diverse training data. Smaller clinics or institutions with limited access to high-quality imaging datasets may face challenges in implementing this approach.

Finally, while fiducial-based validation provides strong evidence for the accuracy of the registration, there are instances where fiducial markers may not be implanted due to patient-specific factors or clinical considerations. In such cases, alternative validation methods or non-invasive markers may need to be explored to assess registration accuracy.

Looking ahead, there are several avenues for improving upon this work. First, the accuracy of the U-Net model can be enhanced by training on larger and more diverse datasets. Expanding the training data to include a wider range of anatomical regions and clinical cases would help mitigate the segmentation accuracy challenges and ensure that the PD metric performs consistently across a broader spectrum of patients. Additionally, exploring advanced DL-based registration methods, such as VoxelMorph (Balakrishnan et al 2019), could offer an alternative approach to integrating segmentation and registration in a unified framework. This could potentially streamline the workflow by reducing the reliance on separate segmentation and registration steps while maintaining accuracy and efficiency.

Expanding the application of this hybrid similarity metric to other anatomical regions and cancer types, beyond the pelvis, presents a promising avenue. The adaptability of the PD metric to different imaging modalities, such as MRI or PET, would enhance its clinical utility in diverse settings. Another area of interest would be the on-line implementation of this method in ART workflows, where speed and accuracy are critical for daily adaptive treatments.

Finally, multi-institutional collaborations could play a crucial role in addressing the current limitations of dataset diversity by contributing a wider range of anatomical cases, imaging modalities, and clinical scenarios. For instance, datasets including diverse pelvic anatomies or other treatment sites such as thoracic or head and neck regions would enable broader validation of the PD metric and CycleGAN-generated sCT images. Such datasets would enhance the robustness and generalizability of the proposed approach, paving the way for its clinical adoption.

## 5. Conclusion

This study demonstrated that the use of CycleGAN-generated sCT images to enhance segmentation accuracy significantly improves registration performance in ART, especially in cases where soft-tissue contrast is poor, such as with CBCT images. By combining intensity-based DIR with a structure-guided PD metric, we address the limitations of traditional intensity-based methods. The PD metric, guided by organ contours, provides structural support to the registration process, yielding better alignment of critical structures like the prostate, bladder, and rectum. The improved segmentation quality achieved through the integration of CycleGAN, and U-Net models directly enhances the robustness of the registration process, highlighting the potential of this approach to streamline ART workflows and improve clinical outcomes. Integrating these methods into ART workflows minimizes manual intervention, enhances daily image registration accuracy, and improves dose delivery precision. This approach minimizes unnecessary radiation exposure to healthy tissues, improves clinical outcomes, and streamlines adaptive treatments. Future work should focus on refining segmentation with larger, more diverse datasets and improving generalizability across anatomical sites and patient populations. Expanding this method to other imaging modalities, such as MRI or PET, could further enhance its clinical utility. This hybrid registration approach represents a significant advancement in ART, with strong potential for clinical adoption.

## Acknowledgements

This material is based upon work supported by the National Science Foundation under Grant Nos. 1553436, 1642345 and 1642380 and the National Institutes of Health under NCI R01CA229178.